\documentstyle[11pt,aaspp4]{article}

\tighten

\newcommand{\kms}{$\,{\rm km\,s^{\scriptscriptstyle -1}}$}
\newcommand{\gtsim}{\ {\raise-0.5ex\hbox{$\buildrel>\over\sim$}}\ }
\newcommand{\ltsim}{\ {\raise-0.5ex\hbox{$\buildrel<\over\sim$}}\ }
\def\Msun{\hbox{$\thinspace M_{\odot}$}}

\def\simlt{\lower.5ex\hbox{$\; \buildrel < \over \sim \;$}}
\def\simgt{\lower.5ex\hbox{$\; \buildrel > \over \sim \;$}}

\journalid{}{}
\articleid{11}{14}

\begin{document}

\title{Keck Spectroscopy of Objects with Lens-like Morphologies in the Hubble
Deep Field\altaffilmark{1,}\altaffilmark{2}}

\author{Stephen E. Zepf\altaffilmark{3}, Leonidas A. Moustakas, and Marc Davis}
\affil{Department of Astronomy, University of California, Berkeley, CA 94720 \\
e-mail: szepf, lmoustakas, mdavis@astro.berkeley.edu}

\vskip 24pt

\altaffiltext{1}{Based on observations obtained at the W.M. Keck Observatory,
which is operated jointly by the California Institute of Technology and
the University of California}
\altaffiltext{2}{Based on observations with the NASA/ESA Hubble Space Telescope,
obtained at the Space Telescope Science Institute, which is operated by 
AURA, Inc., under NASA contract NAS 5-26555}
\altaffiltext{3}{Hubble Fellow}

\begin{abstract}
 
          We present spectroscopy from the Keck telescope of three sets
of objects in the Hubble Deep Field which have lens-like morphologies.
In the case of J123641+621204, which is composed of four objects with
similar colors and a mean separation of \ltsim 0\farcs8, we find at least
two distinct components at redshifts of $z=3.209$ and $z=3.220$ which 
are separated by $0\farcs5$ spatially. Each of these components has 
narrow Ly$\alpha$ emission, and possibly NV emission and SiIV and CIV
in absorption or with a P-Cygni profile.
The second case is J123652+621227, which has an arc-like feature offset
by $1\farcs8$ to the southwest of a red elliptical-like galaxy, and 
a ``counterimage'' offset $1\farcs4$ on the opposite side. We 
tentatively find a 
single line at 5301\AA\ at the spatial position of the counterimage, 
and {\it no} corresponding emission line at the position of the arc. 
The colors of the counterimage are consistent with the identification
of this line as Ly$\alpha$ at $z=3.36$.
The colors of the arc are different than those of the counterimage, 
and thus both the colors and spectra indicate that this object is 
unlikely to be a gravitational lens. 
For a third lensing candidate (J123656+621221), which 
is a blue arc offset by $0\farcs9$ from a red, elliptical-like galaxy, 
our spectroscopy does not clearly resolve the system spatially, 
complicating the interpretation of the spectrum. We discuss possible 
identifications of a number of absorption features and a very tentative 
detection of a pair of emission lines at 5650\AA\ and 5664\AA\/, and 
find that gravitational lensing remains a possibility in this case. 
We conclude that the frequency
of strong gravitational lensing by galaxies in the HDF appears
to be very low.
This result is difficult to reconcile with the introduction of a cosmological
constant to account for the large number of faint blue galaxies via
a large volume element at high redshift, and tends to favor models
in which very faint galaxies are at fairly modest redshifts.

\end{abstract}


\newpage

\section{Introduction}

        Gravitational lensing by galaxies is of interest both because
it provides constraints on the mass distribution of the lensing galaxy,
and because it allows the study of galaxies at higher redshifts than
would otherwise be possible because of the magnification of the lensed
object. The Hubble Space Telescope (HST) has proven to be very useful for
gravitational lensing studies because of its ability to obtain high
resolution images of faint objects. This utility has been demonstrated
both by studies of previously identified lensing candidates
(e.g.\ FSC 10214+4724, Eisenhardt et al.\ 1996), and the discovery of new
candidate lenses (e.g. Ratnatunga et al.\ 1995).

        The HST images of the Hubble Deep Field (HDF) provide a rich
ground for gravitational lens searches because of the combination of
deep exposures in multiple colors and the high spatial resolution of 
the HST (Williams et al.\ 1996). A visual
inspection of the field reveals several possible candidate lensing
systems (e.g.\ Hogg et al.\ 1996). These include a configuration of 
four compact galaxies with similar blue colors, separated by 
$\ltsim 1\farcs0$ (J123641+621204, or in our notation L4.1), 
a red ``elliptical'' with a blue ``arc'' offset by about $1\farcs8$ 
to the southwest and a potential counterimage on the other side 
(J123652+621227, or L3.1), and a red ``elliptical'' with a
blue ``arc'' offset by about $0\farcs9$ to the south (J123656+621221, or L3.2).

        Spectroscopy of the constituent objects is critical for
determining the nature of these candidate gravitational lens
systems. Therefore, as part of a modest program of spectroscopy
of galaxies in the HDF, we used the Keck telescope to obtain low-resolution
spectra of objects in these candidate lenses with the aim of determining
their redshifts and spectral characteristics. The observations
themselves are described in $\S$ 2, the results of the analysis of the
data in $\S$ 3, and the implications of these results are discussed in $\S$ 4.

\section {Observations}

        Spectra of objects in the HDF were obtained on 1996 March 15 (UT)
with the Keck 10-m telescope and the Low-Resolution Imaging Spectrograph
(LRIS, Oke et al.\ 1995). The LRIS slit-mask facility allowed 
us to obtain spectra of many objects simultaneously. We designed 
two slitmasks, each of which contained about 20 objects in the HDF itself 
and its flanking fields. The slitlets were $0\farcs7$ wide. The spectra 
were obtained using the 300 l/mm grating, which provided wavelength 
coverage from about 5000\AA\ through 9500\AA\, with the exact coverage
dependent on the location of the slitlet, and with declining sensitivity 
at the red end of this range.
With the 24$\mu$m pixels of the Tektronix CCD, the wavelength scale is 
approximately 2.4\AA$/$pixel and the spatial scale is $0\farcs214/$pixel.

        In selecting our objects, we specifically targeted the three sets
of objects in the HDF which we found to be the most likely lensed systems.
True-color images of these sets of objects and the orientation of our
slit positions is shown in Figure 1 (Plate 1).
The positions, magnitudes, and colors of the objects in these candidate
lens systems are given in Table 1.
The position angles of the two masks were chosen so that the slitlets
passed through
both the candidate lens and lensed object for the two systems in
which both are identifiable. This corresponds to a position angle
of $19^{\circ}$ (PA19) for L3.2 and $61^{\circ}$ (PA61) for L3.1.
A slitlet was placed across the system of four similar objects
(L4.1) at both position angles.

        Our observations consist of 80 minutes of integration
time with the PA61 mask and 60 minutes for the PA19 mask, each of which
was divided into two exposures.
From images taken during the alignment procedure, we estimate the
seeing was about $0\farcs9$.
The reduction of these data is described in detail in
Moustakas et al. (1996). Briefly, after standard CCD reduction techniques,
the two exposures were combined using bad pixel maps to eliminate cosmic
rays. The combined image was checked by eye for the few cosmic rays which
escaped detection in the first pass. The combined image was then transformed
to linear wavelength scale using arc lamps observed after each exposure.


One-dimensional spectra were then extracted from the two-dimensional
data. The location on the slitlet of each extraction was based on
the observed location of flux. In all cases, this corresponds to the
expected location of a target object. The one-dimensional spectrum
was smoothed with a gaussian of $\sigma = 3.0$\AA\/, which is the
approximate instrumental resolution. No further smoothing was performed.
A relative flux calibration
of the spectra was made using a spectrophotometric standard observed
with LRIS in long-slit mode on an earlier observing run with a very
similar setup (kindly provided by H. Spinrad and D. Stern). Although
this is clearly far from ideal for flux calibration, the result is
consistent with a previously obtained spectrum of one of our targets.
The resulting one-dimensional spectra are described
in detail in the following section.

\section {Analysis}

\subsection {L4.1 (J123641+621204)}

        The extracted spectra of this set of objects are shown
in Figure 2, along with a spectrum obtained for the same
objects with a wider slit by Steidel et al. (1996, hereafter SGDA).
This figure indicates that our spectra confirm SGDA's general
identification of Ly$\alpha$ emission at $z \simeq 3.22$, and 
possibly their identification of SiIV absorption at the same redshift.
This redshift is consistent with the colors
of the objects given in Table 1, as the redshifted Lyman break
provides a natural explanation for the very red F300W $-$ F450W
color. In fact, the objects were selected by SGDA
to be candidate $z \sim 3$ galaxies on the basis of their colors.

	A closer look at our spectra shows that at both of our position
angles, the emission lines have two distinct components. Based on
Ly$\alpha$, the redshifts of these components are $z = 3.220$ 
and $z=3.209$, corresponding to a rest frame velocity difference 
of 800 \kms.  Moreover, our two-dimensional spectra indicate that
the different velocity components are at different spatial locations. 
This is shown in Figure 3 (Plate 2), in which the Ly$\alpha$ lines are 
at the same location or only slightly offset along the slit for 
PA61, but are separated by several pixels for PA19. 
A correlation between velocity and position had been tentatively suggested
by SGDA.
{\it The detection of a different spatial
location for the different velocity components rules out 
gravitational lensing as the cause of this compact configuration
of high redshift objects}. 

	With our two position angles, we can attempt to identify
which of the four objects is responsible for each feature.
Because the emission lines are at roughly the same position on the
PA61 slitlet, the likely candidates are the northern and southern objects, 
as these are the only two at roughly the same position along the slit 
at this angle. This tentative identification appears to be confirmed by
the offset observed on the PA19 slitlet, which is exactly that
expected if the two emission components are from these two objects.
However, as a caution we note the seeing was only marginal for
resolving this system, and a firm identification of individual
spatial components with velocity components awaits spectral data
with better spatial information. 

	We also note that both redshift components appear to have 
NV emission which is strong relative to Ly$\alpha$, and which is redshifted
compared to its expected location given the $z$ determined from Ly$\alpha$.
Strong NV emission is also present in SGDA's spectrum, as is the
offset of the peak emission to slightly higher redshift than that
of the corresponding Ly$\alpha$ feature.
The relative strength of the high ionization NV line is 
suggestive of an AGN (e.g.\ Kinney et al. 1993). 
However, the redward peak of the NV feature and the
possible evidence for some P-Cygni absorption blueward of the expected
line center suggest that there may also be a massive star component 
to the NV emission. Therefore, this system appears to be composed
of two or more star-forming galaxies, possibly with AGN, separated 
by roughly 10 kpc and 800 \kms.

\subsection{L3.1 (J123652+621227)}

	From west to east, our slitlet for this system crossed the 
``arc'', the central ``elliptical'' and then the ``counterimage''
(see Figure 1). An emission line at $5301$\AA\/ at
the position of the counterimage is the only emission feature
detected along this slitlet. A one-dimensional spectrum
extracted at the position of the counterimage is shown in Figure 4.
The two-dimensional spectrum in the wavelength region around 
the detected emission line is shown in Figure 3 (Plate 2).
The emission feature is visible on each separate exposure.
A faint red continuum at the position of the elliptical 
can be seen on the two-dimensional spectrum, but the signal
is insufficient for any analysis. An even weaker blue continuum
is visible on the two-dimensional image at the position of the
arc. Most importantly, no emission line is seen in the arc spectrum 
in the range from $5000$\AA\ to $7200$\AA\/. 

	The presence of an emission line at the position of the
counterimage and the absence of any corresponding line at the
position of the arc is evidence against a lensing origin for
this system. However, the emission feature is not overwhelmingly
strong, so it is interesting to consider other data. The colors 
of the objects given in Table 1 provide independent evidence 
that the arc feature and counterimage are not lensed images
of a single object. Specifically, the counterimage has a 
significantly bluer F450W$-$F606W color than the arc. 
Although slightly different colors are possible for different
arc features (Hogg et al.\ 1996), the combination of the colors 
and our spectral data favor the more straightforward interpretation 
that the arc and counterimage are different galaxies.

	The colors can also help in the identification of the single 
emission line of the counterimage. The colors show that the counterimage
has a flat spectrum from F814W through F450W, and then probably reddens 
into the F300W bandpass. The flat spectral shape is characteristic of
the star-forming galaxies, and the red F300W$-$F450W color, if real,
suggests that $z \gtsim 2.5$, as a result of the Lyman break redshifting 
through the F300W bandpass (e.g.\ SDGA). 
Unfortunately, the faintness of the counterimage means 
that the absence of a detection only requires F300W$-$F450W $> 0.19$ 
at the $3\sigma$ level, and F300W$-$F450W $> 1.39$ at the $1\sigma$ level.
The colors do not clearly establish the presence or absence of a Lyman
break. However, it is difficult to find examples of lower redshift 
galaxies with flat spectra from F814W through F450W and a significantly
redder F300W$-$F450W color.

	Guided by the colors, a natural identification for the single
strong emission line is Ly$\alpha$. This identification places the object
at $z=3.36$. We find no other emission lines in the spectrum, which may
be taken as support for this identification, since we might expect to 
see [OIII]~4959, 5007 and H$\beta$ lines if the observed line at 5301\AA\ is
identified as [OII]~3727 . If the Ly$\alpha$ identification is correct,
then the red magnitude of AB$_{606+814} = 27.2$ corresponds to a 
rest frame luminosity of about $L_{1600} = 4.8 \times 10^{39}~
{\rm ergs}~ {\rm s^{-1}} {\rm \AA^{-1}}$ for $q_{0} = 0.5$,
and $1.5 \times 10^{40}~
{\rm ergs}~ {\rm s^{-1}} {\rm \AA^{-1}}$ for $q_{0} = 0.05$,
with H$_{0} = 75$  \kms Mpc$^{-1}$ in both cases.
For comparison, a star formation rate of 1 \Msun ${\rm yr}^{-1}$
with a Salpeter IMF gives produces roughly $L_{1600} = 10^{40}~ 
{\rm ergs}~ {\rm s^{-1}} {\rm \AA^{-1}}$ (e.g. Leitherer, Robert, \& Heckman
1995). 

\subsection {L3.2 (J123656+6212210)}

	The extracted spectrum of this pair of objects is shown
in Figure 5. Because the separation between the red ``elliptical'' 
and the blue ``arc'' is only about $0\farcs9$, the objects are not 
clearly spatially resolved in our two-dimensional spectrum. Therefore,
the extracted spectrum plotted in Figure 5 is a composite of both
objects. The composite nature of this spectrum complicates the
identification of the lines and the determination of the redshift
of the objects.

	In the one-dimensional spectrum, we tentatively identify a 
number of absorption features and a pair of emission lines at
5650\AA\, and 5664\AA\/. It is natural to associate the absorption 
features and most of the continuum with the elliptical, because
of its red colors and brighter magnitude over the range of wavelengths
covered by our spectrum. Similarly, it is natural to identify the
possible emission lines with the arc, which has blue colors indicative 
of a nearly flat spectrum ($f_{\nu} \propto \nu^{0}$) across the range 
of bandpasses. The continuation of the flat spectrum through the
F300W bandpass also puts an upper limit of about 2.5 on the redshift
of the arc, because of the absence of a Lyman break. Moreover,
the absence of any strong single emission line in such a blue
spectrum (indicative of a starburst or nuclear activity), suggests
that the redshift is greater than about one, or we would expect to
see lines from either [OII] or [OIII] and H$\beta$. 
Thus, the combination of the colors and the spectrum suggests that
the arc is at $1.0 \ltsim z \ltsim 2.5$. Given this range, our most
probable identification of the pair of emission lines is the MgII 
doublet at 2798\AA\/, which would place the arc at a redshift of 1.02.
This redshift is clearly very tentative.

	Adopting a similar approach of first constraining the range
of reasonable redshifts from the colors, we find that the elliptical 
galaxy has colors which are consistent with those of an elliptical at
a redshift of roughly 0.8. This redshift estimate may not be unique,
as it might also be possible to fit the colors with a redshift around
3.5. However, in the high redshift 
case, the object is much brighter than an ordinary galaxy, while at 
$z = 0.8$, the luminosity of the object is close to $L_{*}$. 
Detailed photometric redshift analyses by other groups also favor
the lower redshift, with Cowie (1996) finding $z \simeq 1.1$ and
the approach of Lanzetta, Yahil, \& Fern\'andez-Soto (1996) 
giving $z \simeq 0.72$ (Yahil 1996). At either
redshift, we have failed to convincingly match the observed absorption 
lines with prominent stellar and interstellar absorption lines.
For the reasons given above, a lower redshift appears to be favored,
but this hypothesis remains unproven by our spectroscopy.
Therefore, lensing remains a viable possibility in this case,
with our favored redshift being slightly greater than one for
the arc and somewhat less than one for the elliptical. However,
further spectroscopy is required to confirm or reject this hypothesis.

\section{Discussion}

	Our primary result is that gravitational lensing is unlikely
to be the cause of the configuration of two of the three candidate 
lensing systems we have studied. Specifically, we find that the system 
L4.1 has two distinct velocity components at slightly different 
positions. We have also found some evidence against gravitational 
lensing in the L3.1 system, which was identified by Hogg et al.\ (1996) 
as the most probable candidate for lensing in the HDF.
In the case of the L3.2 system our data are consistent with lensing, 
but are inconclusive.  It is therefore plausible that there 
is no more than one multiply imaging lensing system among the 
approximately 750 objects in the HDF to a limiting magnitude of roughly 
F814W$_{AB} < 27$ for the observed magnitudes of the sources.
Although we have not obtained spectra for every object in the HDF, 
we targeted the three systems which appeared to us to be the most
likely lensing cases, and two of these three have been similarly 
identified by other groups (e.g.\ Hogg et al.\ 1996).
Therefore, we consider it unlikely that 
there is a large population of undiscovered multiple lensing systems
within the magnitude limit given above.

	We can estimate the observed frequency of lensing by combining 
the number of lenses determined above with the number of potential
sources in the HDF. As a starting point we adopt the number of objects
with F814W$_{AB} < 27$, which is similar to the magnitude of the fainter
candidates we studied. This limit gives roughly 750 objects which
are potential lens sources. The true number of potential sources is
likely to be greater, since this number does not account for 
magnification of fainter galaxies.
Taking our observational limit of one lensing system, we derive
a rough lensing rate of around 0.000-0.002; this 
estimate clearly has large uncertainties due to the small
numbers involved. 

	Even with the large uncertainties, it is difficult
to reconcile these observations with the high lensing rates
expected in some cosmological models. In particular, cosmological
models with large $\Lambda$ in which the large numbers of faint 
blue galaxies are a normal evolving galaxy population
appear to be inconsistent with the low rate of multiply
imaging lensing of very faint galaxies.
Although a detailed model of lensing of galaxies in the HDF is not 
yet available, it is notable that the frequency of multiply imaging 
lensing derived above is about an order of magnitude lower than that 
observed for quasars, which itself places tight constraints on the 
cosmological constant (e.g.\ Kochanek 1996 and references therein).
Therefore, a high $\Lambda$, passive evolution model for faint galaxies 
is only viable if there is some combination of effects which reduce 
the observed frequency of galaxy lensing by more than an order of 
magnitude compared to quasars. A factor of several difference is
expected from the lower magnification bias likely for galaxies
relative to quasars. However, the faint galaxies are observed to be 
compact and are expected in these models to be at rather high redshift,
both of which make it hard to reduce the observed 
lensing rate relative to quasars by more than an order of magnitude.
With better 
statistics, it may be possible to constrain other models for the faint
blue galaxies. For example, if the lensing rate for the faint
blue galaxies can be shown to be less than that for quasars and radio
sources, the simplest explanation would be that the faint blue galaxies
are at modest redshifts compared to these other source populations.

\acknowledgments

This work is based in part on observations obtained at the W.M. Keck
Observatory, which is operated jointly by the California Institute
of Technology and the University of California, and would not have been
possible without the Keck telescope and the staff who operate it.
We also acknowledge Bob Williams and the HDF team for carrying out
the survey and making the data immediately available.
We thank Chuck Steidel for providing his spectrum of J123641+621204 in
advance of publication, and Drew Phillips for providing his software
for mask design and alignment, both of which were very useful. Aaron Barth,
Joe Silk, and Hy Spinrad, provided useful comments during the progress 
of this work, and Len Cowie and Amos Yahil provided photometric redshifts
in advance of publication. The suggestions of an anonymous referee improved
the final paper. This research is supported by NSF grant AST92-21540.
S.E.Z. acknowledges support from NASA through grant
number HF-1055.01-93A awarded by the Space Telescope Science Institute,
which is operated by the Association of Universities for Research in
Astronomy, Inc., for NASA under contract NAS5-26555.

\clearpage

 
 
\begin{deluxetable}{lcccc}
\tablecaption{Photometric Data for HDF Objects with Lens-like Morphologies}
\tablehead{
\colhead{Object Name} & \colhead{F814W\tablenotemark{1}} &
\colhead{(F606W - F814W)} &
\colhead{(F450W - F606W)} &
\colhead{(F300W - F450W)\tablenotemark{2}}} 

\startdata
L4.1-N & $25.29$ & $0.19$ & $0.76$ & $2.72$ $(>2.31)$ \nl
\phantom{-------}W & $26.04$ & $0.28$ & $1.02$ & \nodata $(>1.21)$ \nl
\phantom{-------}S & $26.61$ & $0.20$ & $0.95$ & \nodata $(>0.79)$ \nl
\phantom{-------}E & $26.33$ & $0.32$ & $0.92$ & $1.83$ $(>0.98)$ \nl
L3.2-arc & $26.18$ & $0.27$ & $0.05$ & $0.17$ \nl
\phantom{-------}ell & $23.77$ & $1.74$ & $1.88$ & $0.47$ \nl
L3.1-arc\tablenotemark{3} & $26.72$ & $0.41$ & $0.42$ & $1.67$ $(>0.99)$ \nl
\phantom{-------}cnt & $27.18$ & $0.41$ & $0.78$ & \nodata $(>0.19)$  \nl
\phantom{-------}ell & $25.02$ & $1.79$ & $1.68$ & $0.95$ $(>0.06)$ \nl
\tablenotetext{1}{AB magnitude within $0\farcs2$ radius aperture.}
\tablenotetext{2}{If the counts within the $0\farcs2$ radius aperture
are less than $3\sigma$ above the background fluctuations within the 
same aperture, the color at the $3\sigma$ limit is given in parentheses.
No data indicates that the counts are below $1\sigma$ of the background.}
\tablenotetext{3}{Aperture centered at the middle of the primary arc.}
\enddata
\end{deluxetable}


\clearpage
 
\centerline {\bf Figure Captions}

\noindent Figure 1 - True-color images of the lens-like mophology
objects, in which the blue plane is from the F450W image, the green
from the F606W image, and the red from the F814W image.  Each image 
is oriented with North up and East to the left, and is $8\farcs0$ on a
side. The white lines indicate the position angles of the slits.
The image on the top left is L3.1 (J123652+621227), on the top right
is L3.2 (J123656+621221), and on the bottom is L4.1 (J123641+622104).

\noindent Figure 2 - The extracted spectra for L4.1 (J123641+622104).
The two panels are our spectra at the position angles shown, and the
bottom panel is the spectrum from SGDA. All of these show Ly$\alpha$ and
NV emission, and SiIV absorption. Both the Ly$\alpha$
and NV emission lines are double-peaked, and the NV feature is redshifted
relative to Ly$\alpha$ in all of the spectra. The higher resolution
of our spectra relative to SGDA's spectrum results primarily from our 
narrower slit.

\noindent Figure 3 - The top two images are two-dimensional spectra of 
the L4.1 system (J123641+622104) at PA=19$^{\circ}$ and PA=61$^{\circ}$
respectively. These show a spatial shift of about $0\farcs5$ at
PA=19$^{\circ}$ and little or no shift at PA=61$^{\circ}$. 
The lower image is the two-dimensional spectrum of the L3.1 
(J123652+621227) system. The images have been smoothed by an elliptical
Gaussian with a FWHM of the instrumental resolution in the
wavelength direction and the seeing in the spatial direction.
The width of this smoothing is noted in the upper left corner
of each image.

\noindent Figure 4 - The extracted spectra for system L3.1 (J123652+621227).
The top panel shows the spectrum at the position of the counterimage,
and the bottom panel shows the spectrum at the position of the arc-like
feature. The emission line at 5301\AA\ in the spectrum of the
counter-image is not present in the spectrum at the position of the
arc, suggesting that this system is not a gravitational lens.

\noindent Figure 5 - The extracted spectrum of system L3.2 (J123656+621221).
The spectrum is a composite of the elliptical and the arc, as these
were not clearly resolved spatially in our two-dimensional spectrum.
A tentative pair of emission lines and a number of likely
absorption features are marked.

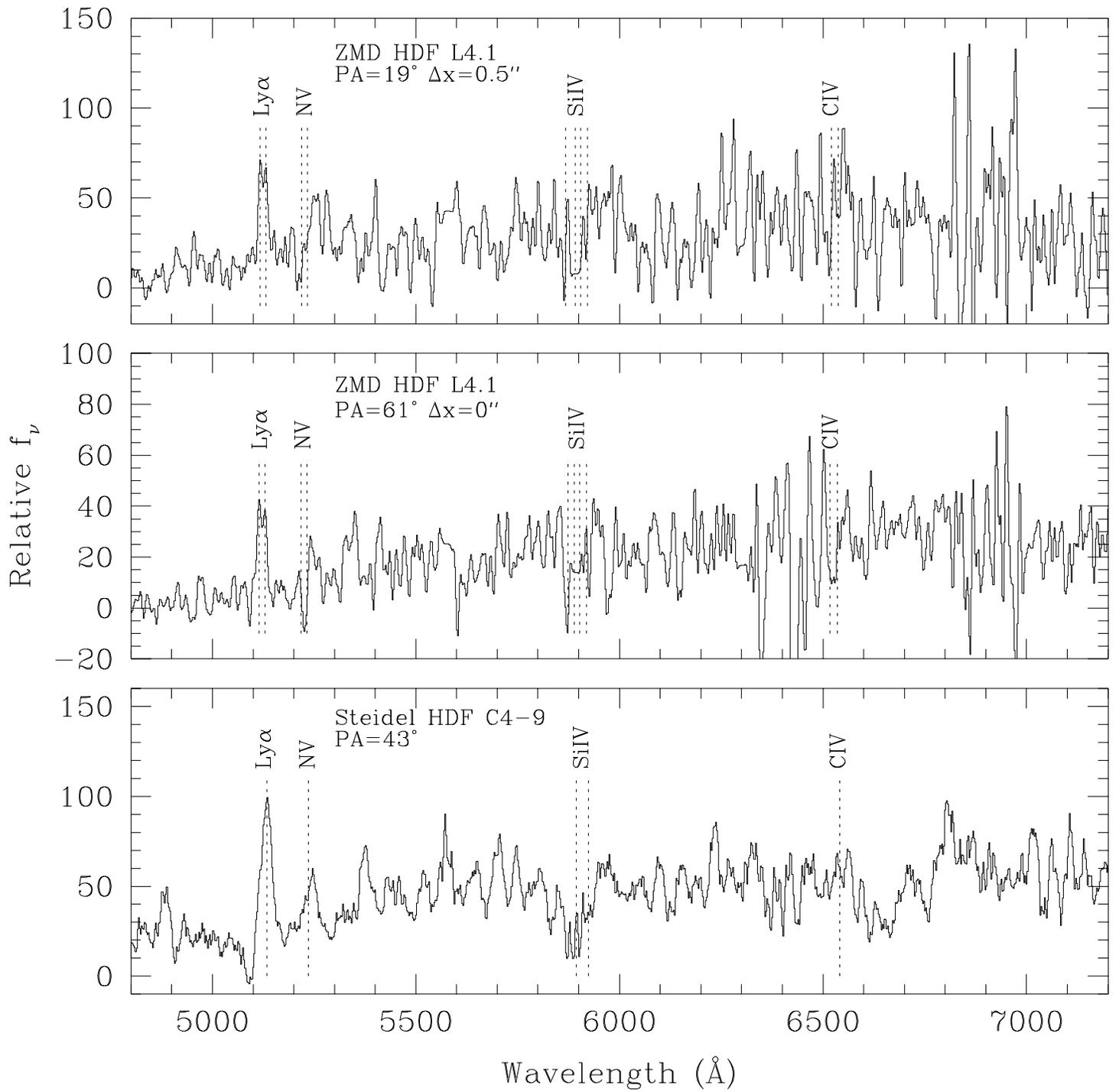
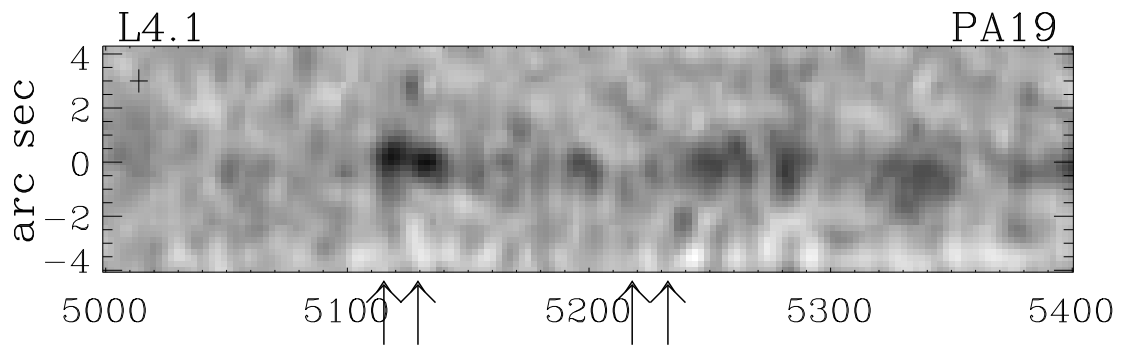
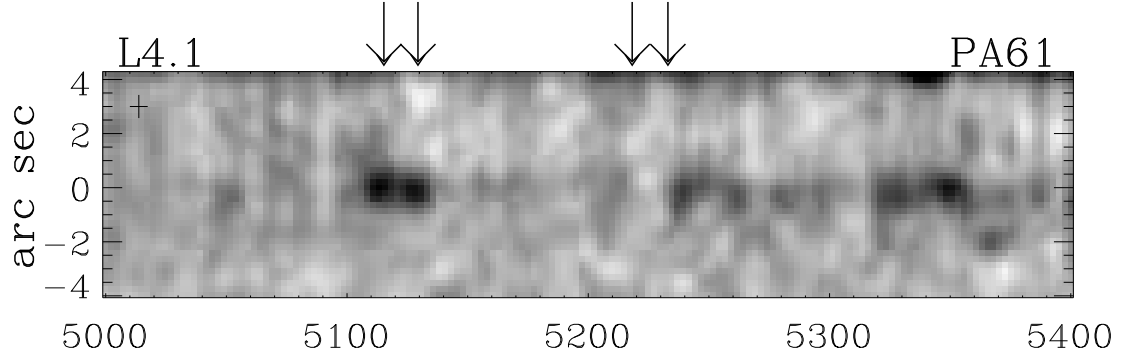
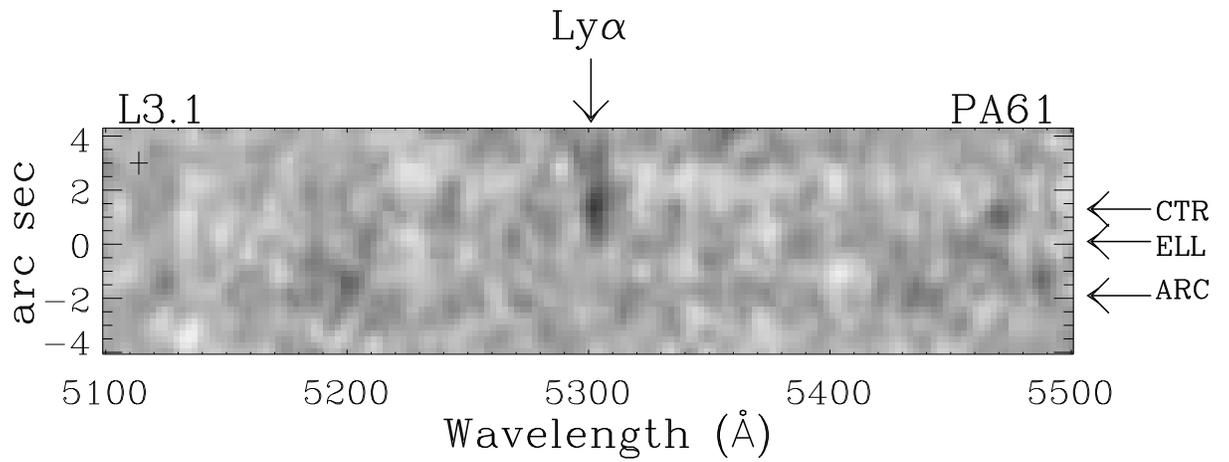
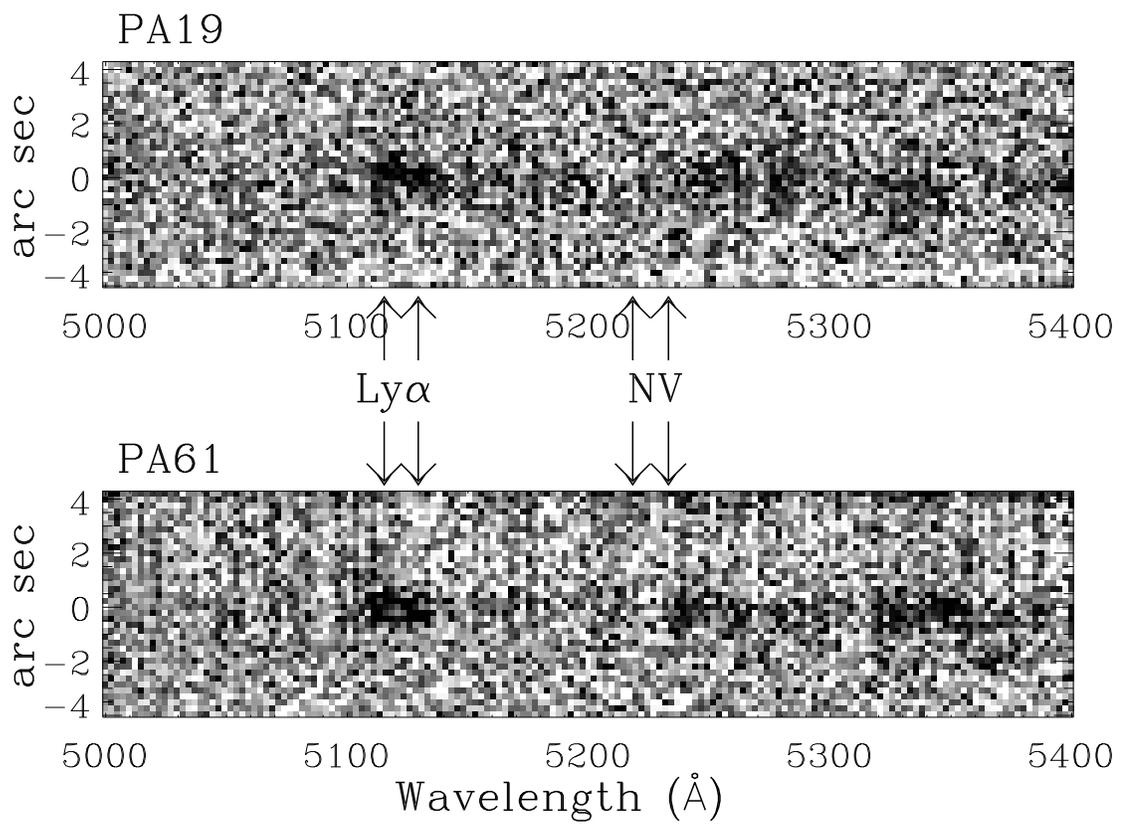
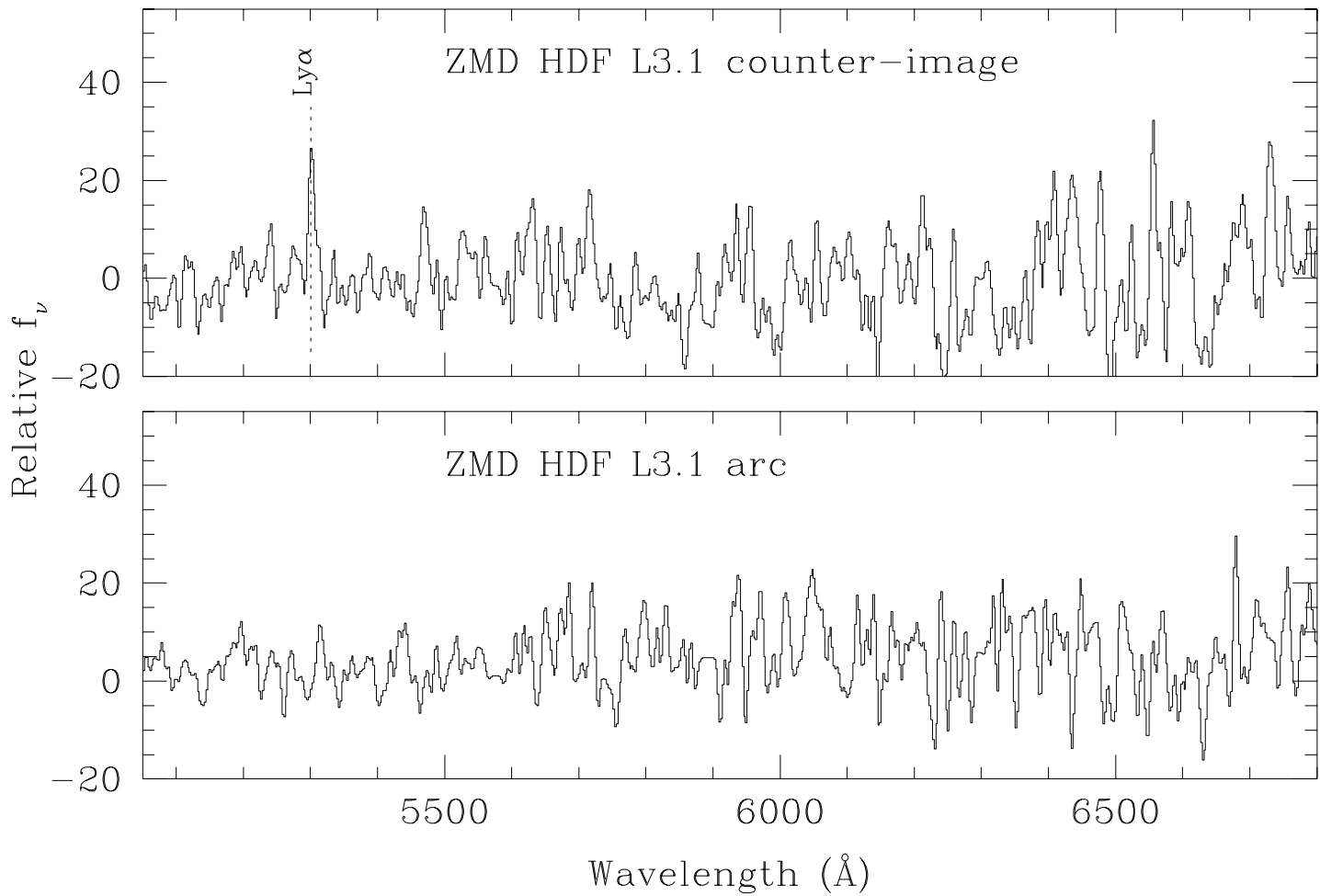
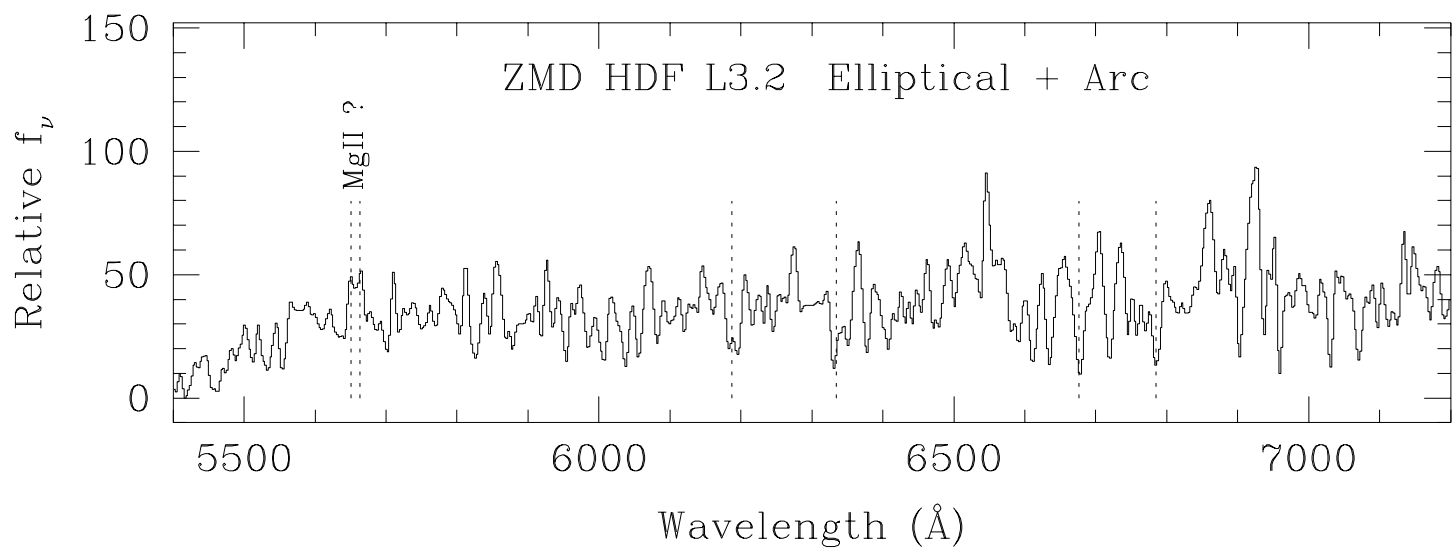
\end{document}